\documentclass[twocolumn]{aastex7}
\usepackage{amssymb,amsmath,}
\usepackage{graphicx}
\usepackage{xcolor,natbib}
\usepackage{multirow}
\usepackage[T1]{fontenc}
\usepackage{ae,aecompl}
\usepackage{newtxtext, newtxmath}
\usepackage{float}
\usepackage{subfigure}
\usepackage{longtable}
\usepackage{enumitem}
\usepackage{longtable,listings}
\usepackage[flushleft]{threeparttable}
\usepackage{parcolumns}
\usepackage{hyperref}

\submitjournal{ApJ}

\shorttitle{AT 2020vdq}
\shortauthors{Zhang}

\begin{document}

\title{Double tidal disruption events or repeating partial tidal disruption events in AT 2020vdq}

\correspondingauthor{XueGuang Zhang}
\author{XueGuang Zhang$^{*}$}
\affiliation{Guangxi Key Laboratory for Relativistic Astrophysics, School of Physical Science and Technology, GuangXi 
University, No. 100, Daxue Road, Nanning, 530004, P. R. China}
\email[show]{xgzhang@gxu.edu.cn}

\begin{abstract}
AT 2020vdq has been known as a candidate of repeating partial tidal disruption events (pTDEs), due to its two flares with a time 
interval of $\sim$1000 days. Here, a simplified method is proposed to test such repeating pTDEs scenario considering 
a main-sequence star tidally disrupted twice. For the two flares in AT 2020vdq if related to the repeating pTDEs scenario, 
theoretical TDE model determined stellar mass of the original star disrupted for the first flare should be not very different 
from the mass of the star (to trace the reminder of the original star) disrupted for the second flare, because a partial TDE with 
impact parameter $\beta$ smaller than 1 can lead to most of (probable higher than 90\%) the stellar mass also bound to the reminder 
of the original star. After considering theoretical TDE model applied to describe the two flares in AT 2020vdq, the model determined 
stellar masses are about 2${\rm M_\odot}$ and $0.36{\rm M_\odot}$ for the stars disrupted in the first flare and the second flare. 
The large mass difference cannot be reasonably expected by the repeating pTDEs with $\beta$ around 0.6 in AT 2020vdq. The results 
in this manuscript indicate that the repeating pTDEs scenario is not preferred at current stage in AT 2020vdq, but the probable 
double TDEs for two individual stars tidally disrupted should be currently recommended.
\end{abstract}


\section{Introduction}

	AT 2020vdq (RA=152.22273, DEC=42.71675, at $z=0.045$) with a re-brightened flare about 1000 days after the first flare has 
been reported as a preferred candidate of repeating Partial Tidal Disruption Events (pTDEs) as recently discussed in \citet{sr25}, 
after considering the re-disrupted reminder of the first disruption of a star. Such similar repeating pTDEs with re-brightened 
features (or repeating flares) have been reported in some known TDE candidates, such as the results in ASASSN-14ko in \citet{ps21, 
ps23}, in eRASSt J045650.3-203750 in \citet{lr23, lr24}, in AT 2020dbl in \citet{lw24}, in AT 2021aeuk in \citet{bd24, sj25}, etc.

	As discussed in \citet{gr13, gs19, lc20, ry20, sp24}, when impact parameter $\beta$ is smaller than the critical value 
around 2, a star around a central supermassive black hole (SMBH) should be not fully disrupted, but a reminder will 
be left out for the re-disruption leading to the expected re-brightened flares, similar as the results expected by 
the repeating pTDEs. Furthermore, as more recent detailed discussions in \citet{sp24} (see their Fig.~5), when the $\beta$ is 
smaller than 1, the mass of the reminder is only a bit smaller than the stellar mass of the original star tidally disrupted. 
Therefore, under the repeating pTDEs scenario, especially with small $\beta$, once there was a second TDE (for the 
second flare) related to the reminder of the first TDE, a basic result can be expected that the stellar mass of the reminder 
should be not apparently smaller than the stellar mass of the disrupted star in the first TDE (for the first flare). This is the 
starting point of the manuscript.

\begin{figure*}
\centering\includegraphics[width = 18cm,height=20.6cm]{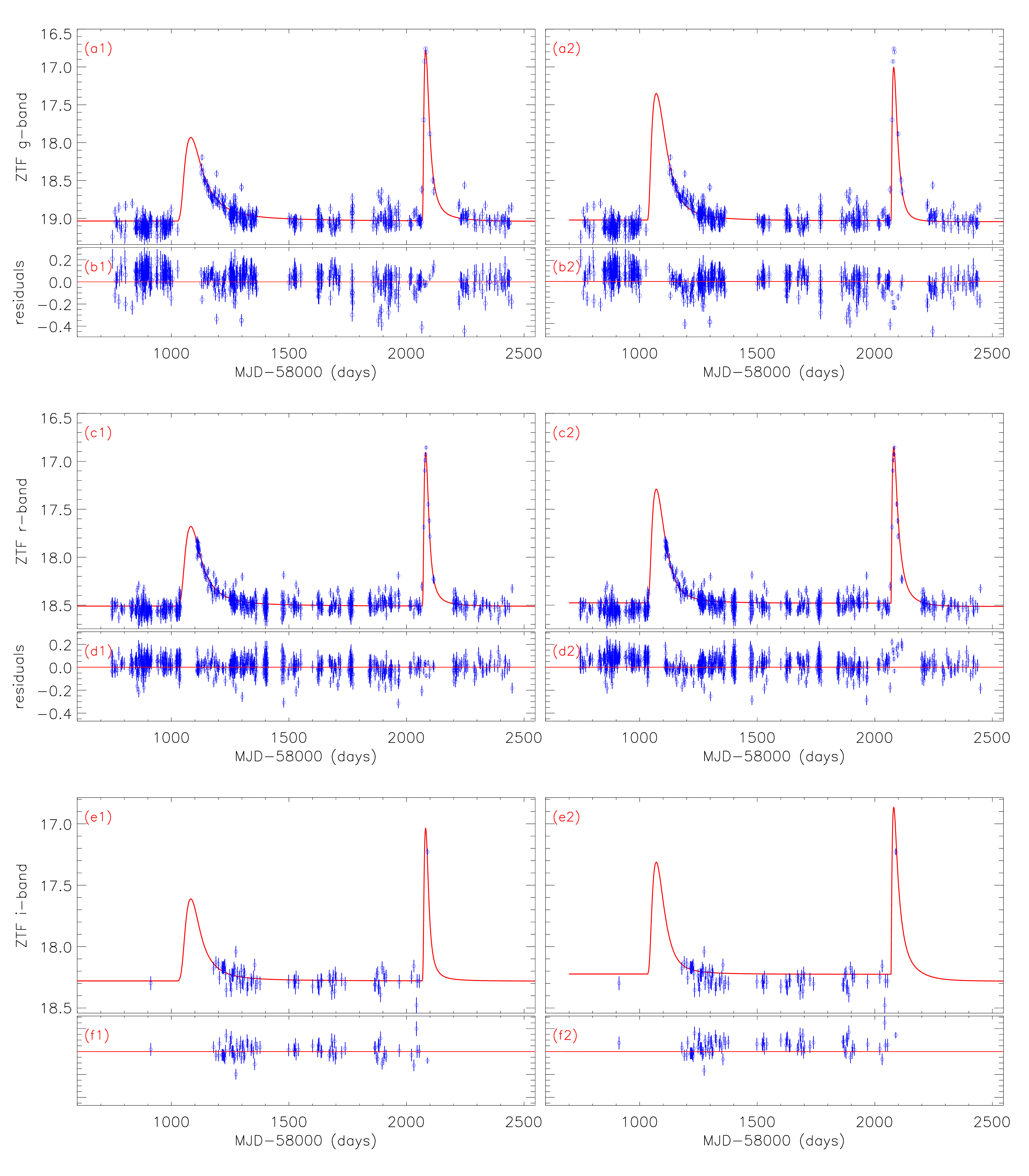}
\caption{Left panels show the best descriptions to the ZTF $gri$-band light curves with different $\beta$ applied to the two 
flares. In left panels, panel (a1), (c1) and (e1) show the ZTF $gri$-band light curves (open circles plus error bars in blue) 
of AT 2020vdq in observer frame, and the corresponding best descriptions (solid red line) by the proposed TDEs model. Panel 
(b1), (d1) and (f1) show the corresponding residuals calculated by the light curves minus the best descriptions, with horizontal 
solid red line as residuals to be zero. Right panels show the best descriptions to the ZTF $gri$-band light curves with the same 
$\beta$ applied to the two flares. In each right panel, symbols and line styles have the same meanings as those in the left panel.}
\label{lmc}
\end{figure*}

	Moreover, as discussed in \citet{sp24}, the reminder of a partially disrupted star can live on the main sequence for 
much longer than the original star they were disrupted from. Therefore, for the re-brightened flare related to the second TDE, 
theoretical TDE model can also be efficiently applied by a main-sequence star tidally disrupted by the central 
SMBH, especially if the time interval is long enough between the two flares. In other words, for two flares related to repeating 
pTDEs scenario, theoretical TDE model applied to describe them can lead to determined stellar masses of the original star and 
its reminder. More specially, if the time interval was long enough between the two flares, indicating few effects of the first 
flare on the following second flare, the theoretical TDE model can be independently applied to determine the stellar masses 
related to the two flares. Then, the mass comparisons between the original star and the reminder could provide further clues 
to support or to be against the repeating pTDEs scenario, which is the main objective of the manuscript.

	The manuscript is organized as follows. Section 2 gives our main results and necessary discussions on variability 
properties of AT 2020vdq through the long-term light curves from the Zwicky Transient Facility (ZTF) \citep{ref8, ref9}, due 
to its long enough time interval between the two flares in AT 2020vdq. Section 3 presents our main summary and conclusions. 
Throughout the manuscript, we have adopted the cosmological parameters of $H_{0}=70{\rm km\cdot s}^{-1}{\rm Mpc}^{-1}$, 
$\Omega_{\Lambda}=0.7$ and $\Omega_{\rm m}=0.3$.

\section{Main results and necessary discussions}

	Detailed descriptions can be found in \citet{sr25} on the long-term variability properties of the 
repeating pTDEs candidate AT 2020vdq. In this manuscript, the ZTF $gri$-band light curves of AT 2020vdq are collected and 
shown in Fig.~\ref{lmc} in observer frame, with apparent first flare around MJD-58000~$\sim$~1100 and the second flare 
around MJD-58000~$\sim$~2100, leading to totally no mixed features between the two flares.

	Theoretical TDE model applied to describe the two flares in AT 2020vdq can be simply done 
as follows. For the first flare described by a standard TDE case, a star with stellar mass $M_{*,1}$ (in units 
of ${\rm M_\odot}$) and stellar radius $R_{*,1}$ (in units of ${\rm R_\odot}$) was disrupted by the central 
SMBH with BH mass $M_{BH6}$ (in units of ${\rm 10^6M_\odot}$). For the second flare described by one another 
standard TDE case, one star (to trace the reminder of the disrupted star in the first TDE case) with stellar mass $M_{*,2}$ 
and stellar radius $R_{*,2}$ was disrupted by the same central SMBH, with the same energy transfer efficiency 
$\eta$ and the same polytropic index $\gamma$ as those applied in the first TDE case.

	Similar as what have recently done in \citet{zh25a} to describe the plateau features in optical light curves of the 
changing-look active galactic nuclei Mrk1018, and also similar as what we have done to describe TDE expected long-term 
variabilities in \citet{ref27, ref28, ref29, ref30, zh25b}, the procedure to describe the two flares in AT 2020vdq can be 
applied by the following three steps.

	First, accepted the theoretical TDE model (the public codes of MOSFIT/TDEFIT) \citep{ref13, ref14} expected fallback 
material rate $\dot{M_{fbt}}(\beta,t)$, through the basic mass conservation assumption as described in \citet{ref14}, the 
corresponding physical viscous-delayed accretion rates $\dot{M}_{a}(t)$ can be determined. Then, for the two TDE 
cases, their physical accretion rates $\dot{M}_1(t)$ and $\dot{M}_2(t)$ can be described as
\begin{equation}
\begin{split}
	\dot{M}_1(t)&=\dot{M}_{a}(M_{BH},~M_{*,1},~T_{vs,1},~\beta_1,~t) \\
	\dot{M}_2(t)&=\dot{M}_{a}(M_{BH},~M_{*,2},~T_{vs,2},~\beta_2,~t)
\end{split}                                                    
\end{equation}
with different viscous delay time $T_{vs,~k}(k=1,~2)$, different impact parameter $\beta_{k}(k=1,~2)$, different stellar 
masses $M_{*,~k}(k=1,2)$ of the two disrupted main-sequence stars. Here, the more recent mass-radius relation in \citet{ek18} 
has been accepted for the main-sequence stars, leading the stellar radius $R_{*,k}(k=1,2)$ not to be a free model parameter 
in our procedure.

	Second, accepted the blackbody photosphere model described in \citet{ref14}, the time dependent output spectrum in 
rest frame can be estimated as 
\begin{equation}
	F_\lambda(t)~=~F_{\lambda,~1}(t)~+~F_{\lambda,~2}(t)
\end{equation}
with $F_{\lambda,~1}(t)$ and $F_{\lambda,~2}(t)$ as the time dependent output spectrum from the first flare 
and the second flare. And $F_{\lambda,~k}(t) (k=1,~2)$ can be described as 
\begin{equation}
\begin{split}
&F_{\lambda,~k}(t)=\frac{2\pi Gc^2}{\lambda^5}\frac{1}{exp(hc/(k\lambda T_{BB,~k}(t)))-1}
	(\frac{R_{BB,~k}(t)}{D})^2 \\
	&R_{BB,~k}(t) = R_{0,k}\times a_p(\frac{\epsilon\dot{M}_{k}(t)c^2}
	{1.3\times10^{38}M_{\rm BH}/{\rm M_\odot}})^{l_{p,k}} \\
	&T_{BB,~k}(t)=(\frac{\epsilon\dot{M}_k(t)}c^2{4\pi\sigma_{SB}R_{BB,~k}^2})^{1/4} \\
&a_{p,~k} = (G M_{\rm BH}\times (\frac{t_{p,~k}}{\pi})^2)^{1/3}
\end{split}
\end{equation}
with $D$ as the distance to the earth determined by the redshift $z$, $k$ as the Boltzmann constant, 
$T_{BB,~k}(t)$ ($k=1,~2$) and $R_{BB,~k}(t)$ ($k=1,~2$) as the time-dependent effective temperature and radius of the 
photosphere related to the corresponding TDE cases, respectively, and $\epsilon$ as the energy transfer 
efficiency smaller than 0.4, $\sigma_{SB}$ as the Stefan-Boltzmann constant, and $t_{p,~k}$ ($k=1,~2$) as time 
information of the peak accretions of $\dot{M}_{k}(t)$ ($k=1,~2$).

	Third, based on the expected emission spectrum $F_\lambda(t)$ in observer frame convoluted with the transmission 
curves of ZTF $gri$ filters, the TDE expected time-dependent ZTF $gri$-band magnitudes can be determined as  
$mag_{T}(g,~r,~i)(t)$. Meanwhile, the host galaxy contributions have been determined as $mag_H(g,~r,~i)=19.04,~18.51,~18.28$ 
by the mean magnitudes of the $gri$-band light curves within MJD-58000 from 1500 to 1900, due to no variability 
during the time duration. Then, the observed time dependent magnitudes $mag_{O}(g,~r,~i)(t)$ (the ZTF $gri$-band 
light curves) can be described by
\begin{equation}
\begin{split}
A~&=~-0.4~\times~mag_O(g,r,i)(t) \\
B~&=~-0.4~\times~mag_{T}(g,~r,~i)(t) \\
C~&=~-0.4~\times~mag_H(g,~r,~i) \\
10^{A}~&=~10^{B}~+~10^{C} 
\end{split}
\end{equation}.

	Based on the described procedure above, there are 12 free model parameters, the central SMBH mass $M_{BH6}$, 
the energy transfer efficiency $\eta$, the stellar mass $M_{*,1}$ of the original star for the first TDE, the impact 
parameter $\beta_1$ and the viscous delay time $T_{vs,~1}$ and $R_{0,1}$ and $l_{p,1}$ for the first TDE case, the 
stellar mass $M_{*,2}$ of the second star (to trace the reminder of the original star) for the second TDE, the impact 
parameter $\beta_2$ and the viscous delay time $T_{vs,~2}$ and $R_{0,2}$ and $l_{p,2}$ for the second TDE case. 

\begin{figure*}
\centering\includegraphics[width = 18cm,height=11cm]{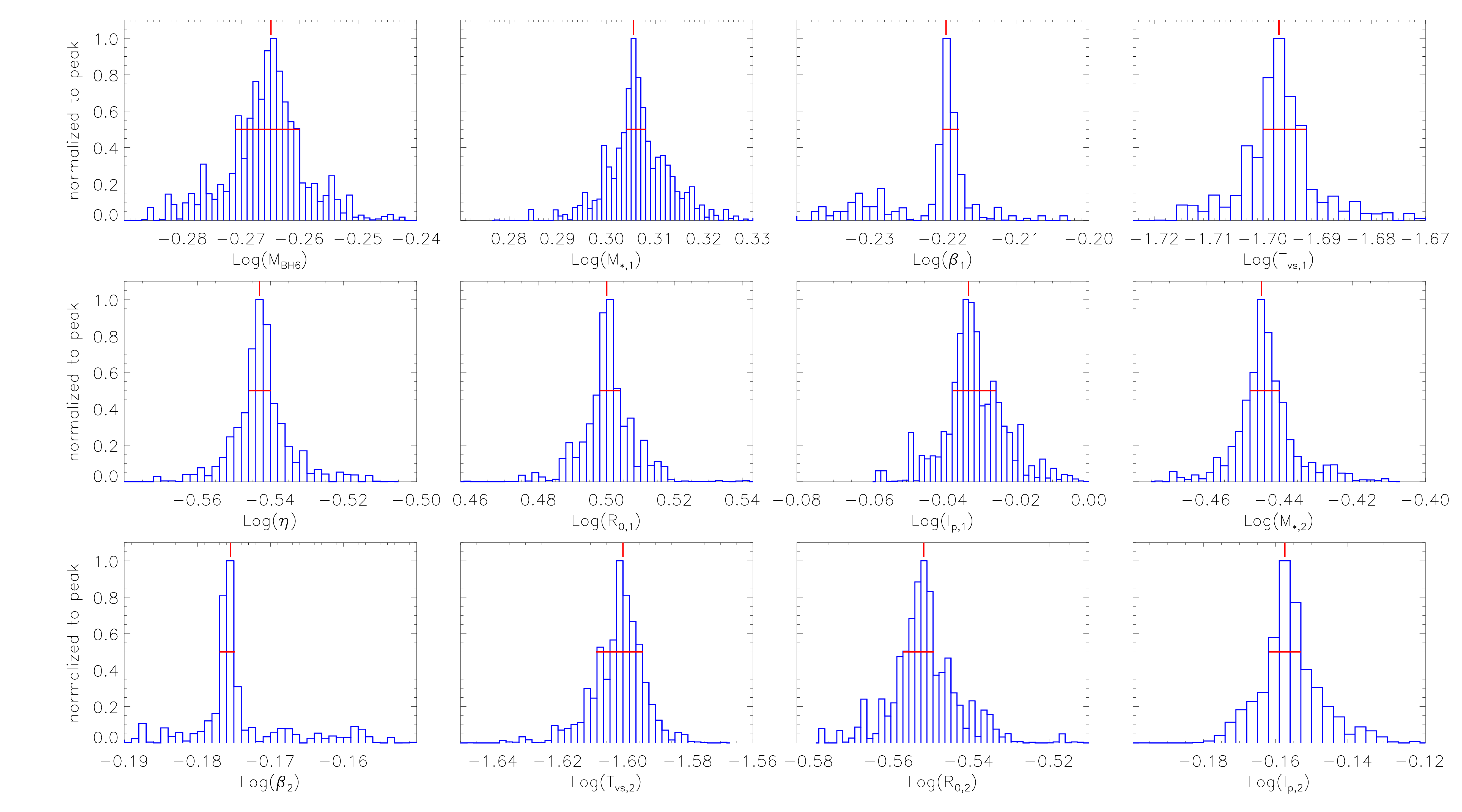}
\caption{The posterior distributions of the model parameters. In each panel, the vertical solid red line marks the 
position of the accepted value of the model parameter, the horizontal solid red line marks the full width at half 
maximum which has been accepted to determine the lower and upper boundaries of the corresponding $1\sigma$ uncertainties 
of the model parameter.}
\label{tde}
\end{figure*}

	Then, through the maximum likelihood method combining with the Markov Chain Monte Carlo (MCMC) technique 
\citep{fh13} applied with the prior uniform distributions of the model parameters as listed in Table~1 (which 
are similar as the prior distributions in \citet{ref14}), the best fitting results can be determined and shown in the 
left panels of Fig.~\ref{lmc} to the ZTF $gri$-band light curves of AT 2020vdq, with determined $\chi^2/dof\sim2389/1264=1.89$. 
The corresponding posterior distributions of the 12 model parameters are shown in Fig.~\ref{tde}, leading to the determined 
robust model parameters and the corresponding 1$\sigma$ uncertainties listed in the fourth column of Table~1. Here, the 
starting values of the model parameters listed in the third column of Table~1 were determined through the Levenberg-Marquardt 
least-squares minimization technique (the known MPFIT package) \citep{mc09}.

\begin{table}[]
\caption{Basic information of model parameters}                
\begin{center}
\begin{tabular}{lllll}     
\hline\hline               
	par & range & value0 & value1 & value2\\   
\hline                      
$M_{BH6}$ & [$10^{-3}$,~$10^3$] & 0.55 & $0.543_{-0.004}^{+0.009}$ & $0.198_{-0.003}^{+0.007}$ \\
$\eta$  &   [$10^{-3}$,~0.4] &  0.29 &  $0.286_{-0.002}^{+0.002}$ & $0.257_{-0.003}^{+0.003}$\\
$M_{*,1}$ & [$10^{-3}$,~50]   & 2.03 & $2.021_{-0.007}^{+0.012}$ & $1.501_{-0.005}^{+0.011}$\\
$M_{*,2}$ & [$10^{-3}$,~50]   &  0.36 & $0.359_{-0.003}^{+0.004}$ & $0.580_{-0.003}^{+0.005}$\\
$\beta_{1}$  & [0.6,~4]    & 0.60 & $0.603_{-0.001}^{+0.003}$ & $0.629_{-0.002}^{+0.003}$\\
$\beta_{2}$  & [0.6,~4]    &  0.67 & $0.667_{-0.002}^{+0.001}$ & $=0.629$ \\
$T_{vs,1}$   & [$10^{-3}$,~3000]    &  7.28 & $7.332_{-0.055}^{+0.086}$ & $17.470_{-0.013}^{+0.017}$\\
$T_{vs,2}$   & [$10^{-3}$,~3000]    &  9.16 & $9.168_{-0.171}^{+0.132}$ & $10.076_{-0.011}^{+0.014}$\\
$R_{0,1}$    & [$10^{-4}$,~$10^4$]  & 3.15 & $3.162_{-0.015}^{+0.029}$ & $0.763_{-0.009}^{+0.009}$\\
$R_{0,2}$    & [$10^{-4}$,~$10^4$]  & 0.28 & $0.281_{-0.003}^{+0.002}$ & $0.604_{-0.008}^{+0.010}$\\
$l_{p,1}$    & [0,~4] & 0.92 & $0.927_{-0.009}^{+0.016}$ & $0.925_{-0.011}^{+0.013}$\\
$l_{p,2}$    & [0, 4] & 0.70 & $0.696_{-0.007}^{+0.007}$ & $0.007_{-0.013}^{+0.017}$\\
\hline                                 
\end{tabular}\\
\end{center}
Note: The first column shows the name of the model parameter. Among the model parameters, the $M_{BH6}$ is in units of 
$10^6{\rm M_\odot}$, the $M_{*,1}$ and $M_{*,2}$ are in units of ${\rm M_\odot}$, the $T_{vs,1}$ and $T_{vs,2}$ are in 
units of days. The second column shows the limited range of the prior uniform distribution of the model parameter. The 
third column shows the starting value of the model parameter. The fourth column shows the MCMC determined accepted value 
of the model parameter and the corresponding $1\sigma$ uncertainties through the applications of TDE model with $\beta$ 
as random parameters for the two flares. The final column shows the determined value of the model parameter and the 
corresponding $1\sigma$ uncertainties through applications of TDE model with the same $\beta$ for the two flares. 
\end{table}

	Based on the determined model parameters, for the first flare in AT 2020vdq, there was a main-sequence 
star (polytropic index $\gamma=4/3$) with stellar mass about 2.02${\rm M_\odot}$ and stellar radius about 2.17${\rm R_\odot}$ 
(determined by the mass-radius relation) tidally disrupted by the central SMBH with BH mass $5.4\times10^5{\rm M_\odot}$, 
leading to the corresponding tidal disruption radius about 60.8$R_G$ ($R_G$ as the Schwarzschild radius of the central SMBH). 
Meanwhile, for the second flare in AT 2020vdq, there was a main-sequence star (polytropic index $\gamma=4/3$, same as the one 
for the star disrupted for the first flare) with stellar mass about 0.36${\rm M_\odot}$ and 
stellar radius about 0.30${\rm R_\odot}$ tidally disrupted by the central SMBH with BH mass $5.4\times10^5{\rm M_\odot}$, 
leading to the corresponding tidal disruption radius about 15.1$R_G$.

\begin{figure}
\centering\includegraphics[width = 9cm,height=6cm]{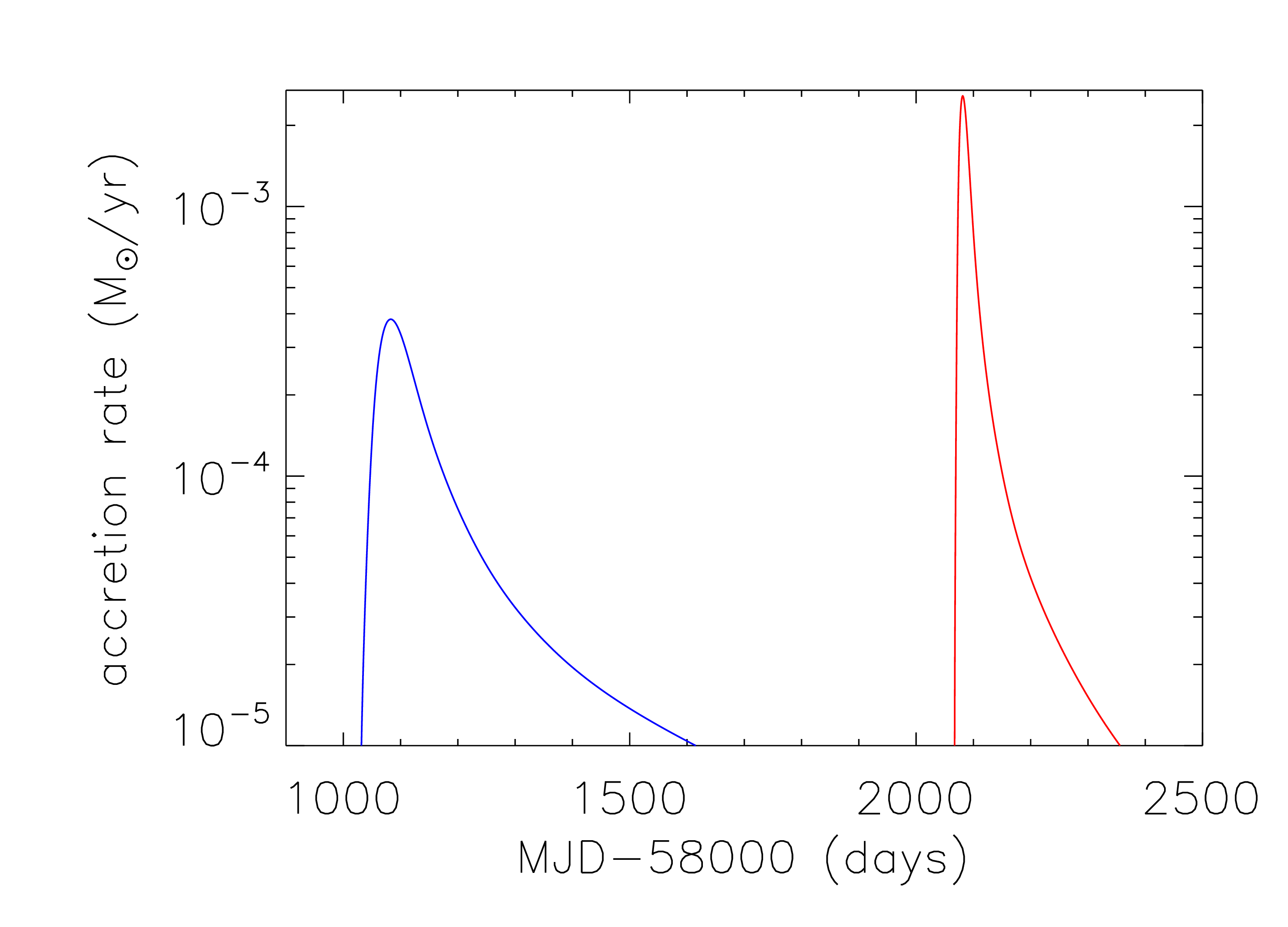}
\caption{Time dependent evolution of the physical accretion rates in AT 2020vdq. Solid line in blue and in red show the 
corresponding results for the first TDE and the second TDE, respectively.}
\label{dm}
\end{figure}

\begin{figure}
\centering\includegraphics[width = 9cm,height=5.5cm]{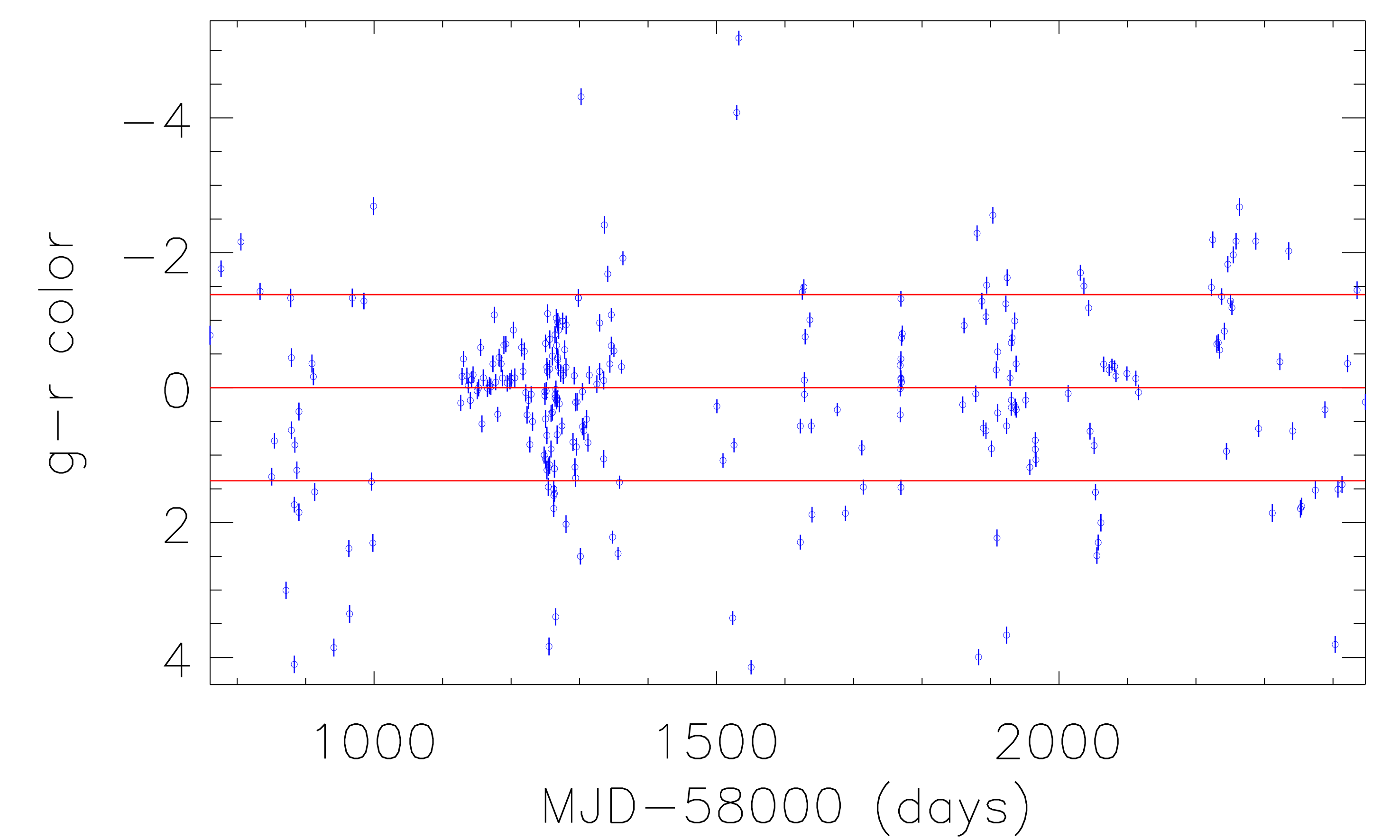}
\caption{The color between ZTF g-band and r-band, after subtractions of host galaxy contributions in AT 2020vdq. Horizontal 
red lines show the color equal to zero and the corresponding 1RMS standard deviations.}
\label{color}
\end{figure}

	Before proceeding further, simple discussions can be given on the mass of the central SMBH in AT 2020vdq. As 
discussed in \citet{yr23}, the BH mass has been estimated to be $3.9_{-1.9}^{+3.7}\times10^5{\rm M_\odot}$ 
through the M-sigma relation \citep{kh13} applied with the measured stellar velocity dispersion of the host galaxy of 
AT 2020vdq. The BH mass determined by the M-sigma relation is consistent with the TDE model determined BH 
mass, providing further clues to support that the theoretical TDE model determined results are reasonable 
in AT 2020vdq. Meanwhile, for the listed determined model parameters in Table~1, comparing the values with 
those of the other reported optical TDEs, such as the ones listed in \citet{ref14}, the values for the model parameters 
of the two flares in AT 2020vdq are common.

	Based on the fitting procedure to describe the flares in AT 2020vdq, the time dependent evolution 
of physical accretion rates $\dot{M_{i}}(t)~(i=1,2)$ can be determined and shown in Fig.~\ref{dm}. We can estimate that 
the total accreted mass is about $1.3\times10^{-4}{\rm M_\odot}$ (corresponding total energy about $6.5\times10^{49}$ergs) 
in the first flare through the determined $\dot{M_1}(t)$ and the energy transfer efficiency $\eta$. In other 
words, about 99.99\% of the total stellar mass of the original star has been restored into the reminder in the first flare 
in AT 2020vdq. Considering the discussed results in \citet{sp24}, for a main-sequence star with stellar mass about 
2${\rm M_\odot}$ tidally disrupted with $\beta\sim0.6$, the mass ratio of the reminder to the original star should be 
higher than 90\% (almost near to 100\%), simply consistent with the determined result in AT 2020vdq.

	After considering the determined mass of the reminder of the disrupted original star in the first flare, it will 
be expected that the reminder with stellar mass about 2${\rm M_\odot}$ (2.02${\rm M_\odot}$-$1.3\times10^{-4}{\rm M_\odot}$) 
should be tidally disrupted for the second flare in AT 2020vdq. Unfortunately, based on the determined model parameters, 
the stellar mass of the star is about 0.36${\rm M_\odot}$ for the second flare, which is about 5.6 times smaller than 
the expected mass of 2${\rm M_\odot}$.

	Besides the accreted mass in the first flare, some unbound materials should be not bound to the central SMBH 
nor bound to the reminder of the original star, leading to the unbound materials falling into the surrounding environments. 
If it was the case  to explain the determined mass difference above, there should be about 
2${\rm M_\odot}$-0.36${\rm M_\odot}$=1.64${\rm M_\odot}$ materials surrounding the central region, which probably lead 
to apparent obscuration. However, after removing the host galaxy contributions, checking the color between ZTF g-band 
and r-band, there are no clues for any obscuration on central flare, as shown in Fig.~\ref{color}. 
Therefore, even considering unbound materials, it is hard to explain the mass difference between the 
reminder expected by the first TDE case and the star expected by the second TDE case in AT 2020vdq.

	Based on the discussed results above, rather than the repeating pTDEs scenario, the double TDEs (including 
two individual TDEs) scenario as discussed in \citet{ml15, zh25a} should be preferred in AT 2020vdq, after 
considering a binary star system tidally disrupted. A binary star system includes the primary star with stellar mass 
about 2.02${\rm M_\odot}$ and the secondary star with stellar mass about 0.36${\rm M_\odot}$. When the 
binary star system was wandering around the central SMBH in AT 2020vdq, followed the primary star partially tidally 
disrupted, then the secondary star was tidally disrupted.

	Furthermore, for the second flare in AT 2020vdq, the total accreted mass is about 
$1.9\times10^{-4}{\rm M_\odot}$ (corresponding total energy about $9.6\times10^{49}$ergs) through the shown 
$\dot{M_2}(t)$ (solid red line) in Fig.~\ref{dm}. Therefore, under the repeating pTDEs scenario, it will be necessary 
to check properties of the next flare in AT 2020vdq. If the next flare in AT 2020vdq can lead the mass of the partially 
disrupted star to be still around 0.36${\rm M_\odot}$, the repeating pTDEs scenario should be strongly supported in 
AT 2020vdq. And the mass difference discussed in the manuscript should be governed by unclear mechanisms. Meanwhile, 
as discussed in the known repeating pTDEs candidate in ASASSN-14ko \citep{ps21, ps23}, if such similar repeating pTDEs 
could be efficient in AT 2020vdq, periodic flares could be expected in AT 2020vdq. Once the time 
interval about 1001days between the peaks of the two flares could be simply accepted as the periodicity of expected 
periodic flares related to repeating pTDEs scenario in AT 2020vdq, the next flare in AT 2020vdq could be 
expected on February 11, 2026. In one word, to test the expected flare after several months and its related physical 
properties could provide direct evidence to support or to be against the repeating pTDEs scenario in 
AT 2020vdq.

	Before ending the section, four additional points should be noted. First, as shown in \citet{ps21, ps23} for 
the repeating pTDEs candidate ASASSN-14ko, the repeating flares have almost similar variability profiles, such as the 
shown stacked phase-folded light curves in Figure 5 in \citet{ps21} and in Figure 3 in \citet{ps23}. However, as shown in 
Fig.~\ref{lmc} for AT 2020vdq, the two flares have very different variability profiles, such as the very 
different peak intensities and very different time durations of the two flares, which could be accepted as indirect 
evidence to probably support that the repeating flares in AT 2020vdq should have some different physical origin for 
the repeating flares in the ASASSN-14ko. Further study on similarity of the variability profiles of 
repeating pTDEs is being prepared and should be reported in the near future.

	Second, there are no discussions on detection rates for double TDEs or repeating pTDEs in this manuscript. 
In other words, we do not to support or to be against the repeating pTDEs scenario by expected detection rates, but 
to check such repeating pTDEs scenario by stellar mass comparisons between the original star and the reminder through 
applications of two individual TDEs cases to describe the variability of the two flares, especially due 
to time interval between the two flares longer enough to apply two theoretical TDE cases in AT 2020vdq.

	Third, considering repeating pTDEs expected totally similar $\beta$ for the tidally star in the first flare 
and the second flare, besides the TDE model determined results above shown in the left panels of Fig.~\ref{lmc}, 
procedure above with the same $\beta$ for the two flares have been applied. The determined results are shown in the 
right panels of Fig.~\ref{lmc} with determined $\chi^2/dof\sim3664/1265=2.90$, and the 
determined values and corresponding uncertainties of the model parameters are listed in the last column of Table~1. 
The very different masses, 1.50${\rm M_\odot}$ and 0.58${\rm M_\odot}$, between the tidally disrupted stars for the 
two flares (corresponding $\beta$ to be 0.629 smaller than 1) can be re-confirmed, indicating the repeating pTDEs 
scenario probably not preferred at current stage. Meanwhile, similar as what we have recently done in \citet{zh22, 
zh23c, zh24c}, through the F-test technique applied with the different $\chi^2$ and $dof$ for the procedures with 
different $\beta$, the results related to the procedure with different $\beta$ for the two flares shown 
in the left panels of Fig.~\ref{lmc} are preferred with confidence level higher than 10$\sigma$. Therefore, there 
are not further discussions any more on the results related to the procedure with the same $\beta$ applied to the 
two flares.

	Fourth, we should note that the discussed results above are mainly through the physical process described 
by theoretical TDE model (MOSFIT/TDEFIT). Once the intrinsic physical process for the two flares in AT 2020vdq 
were very different from the process described by the MOSFIT/TDEFIT, the discussed results should not have deeply 
physical meanings. However, the best fitting results shown in Fig.~\ref{lmc} indicate that the physical process 
described by MOSFIT/TDEFIT can be accepted as a preferred process at current stage to describe the two flares in AT 
2020vdq, unless there were stable evidence to rule out the MOSFIT/TDEFIT described procedure to describe 
the flares in AT 2020vdq. Furthermore, in the process above, distortions of the outer layers of the reminder are 
not considered. Here, effects of the distortions can be simply discussed as follows. The distortions can lead to 
probably steeper dependence $R_*\propto M_{*}^{1/\alpha}$ ($\alpha>1$) (same stellar mass leading to larger stellar 
radius) of stellar radius on stellar mass than the common dependence $R_{*}\propto M_{*}^{\sim1}$ for main-sequence 
stars. Considering the dependence of accretion rate on stellar parameters $\dot{M}\propto M_{*}^{2}R_*^{-3/2}$ as 
shown in equation (4) in \citet{ref14}, for common main-sequence stars, the accretion rate can be simply estimated 
as $\dot{M}_0\propto M_{*}^{\sim1/2}$. However, considering distortions of the outer layers of the reminder should 
lead to $\dot{M}_1\propto M_{*}^{2 - \frac{3}{2\alpha}}$ which is larger than $\dot{M}_0$, indicating smaller stellar 
mass can lead to similar accretion rate. Therefore, considering distortions of the outer layers of the reminder 
should indicate smaller stellar mass of the reminder disrupted for the second flare in AT 2020vdq, leading to larger 
stellar mass difference between the original star and the reminder, to further confirm the mass difference.

\section{Summary and Conclusions}

	Our summary and main conclusions are as follows.
\begin{itemize}
\item A simple method is proposed in this manuscript to check the repeating pTDEs scenario for a main-sequence star 
	tidally disrupted twice in the known AT 2020vdq of which light curves have re-brightened flares.
\item Under the repeating pTDEs scenario, the determined stellar masses for the two flares in turn should be not 
	very different, due to the very low mass loss of the original star disrupted in the first flare, especially 
	for smaller impact parameter.
\item considering the long enough time interval between the two flares in AT 2020vdq, the applications of two 
	individual TDEs to describe the two flares, the stellar mass of the disrupted star in the first TDE case is 
	about 5.5 times larger than the mass of the disrupted star in the second TDE case.
\item Based on the determined model parameters, the total accreted stellar masses are around 
	$10^{-3~\sim~-4}{\rm M_\odot}$ in the two TDE cases for the two flares in AT 2020vdq.
\item The large mass difference cannot be reasonably expected by partially disrupted cases with impact parameter 
	$\beta$ around 0.6, even after considering probable unbound materials falling into surroundings.
\item Therefore, at current stage, rather than the repeating pTDEs scenario for a main-sequence star tidally 
	disrupted twice, the double TDEs with two stars in a binary star system tidally disrupted should 
	be preferred in AT 2020vdq. 
\item If the repeating pTDEs scenario was intrinsically preferred in AT 2020vdq, the next flare should be expected on 
	February 11, 2026. After only several months, there should be clear clues to support or to rule out the 
	repeating pTDEs scenario in AT 2020vdq. 
\end{itemize}

\section*{Acknowledgements}
Zhang gratefully acknowledge the anonymous referee for giving us constructive comments and suggestions to greatly improve 
the paper. Zhang gratefully thanks the grant support from GuangXi University and the grant support from NSFC-12173020 and NSFC-12373014 
and the support from Guangxi Talent Programme (Highland of Innovation Talents). The manuscript has made use of the data from the ZTF 
(\url{https://www.ztf.caltech.edu/}).The paper has made use of the code of TDEFIT/MOSFIT (\url{https://mosfit.readthedocs.io/}) and 
of the emcee (\url{https://emcee.readthedocs.io/}) and of the MPFIT (\url{https://pages.physics.wisc.edu/~craigm/idl/cmpfit.html}).

\end{document}